\documentclass[aps,prb,showpacs,showkeys]{revtex4}
\usepackage[T2A,OT1]{fontenc}
\usepackage[cp1251]{inputenc}         % Кодировка документа
\usepackage[russian,ukrainian,english]{babel} % Многоязычность документа
\usepackage[dvips]{epsfig}            % загрузить пакет epsfig и использовать драйвер dvips
\usepackage{latexsym}
\usepackage{amsfonts}
\usepackage{boxedminipage}

\renewcommand{\theenumi}{\arabic{enumi}}

\begin{document}
\selectlanguage{english} %\sloppy \hyphenation{пру-ж-ні
%влас-ти-вос-ті}
%\title{Pressure and temperature  dependencies of sound velocities and lattice parameters in
%$\beta-$, $\alpha-$ and $\delta$ phases of solid oxygen}
\title{Shift of close-packed basal planes as an order parameter of
transitions between antiferromangetic phases in solid oxygen: II.
Temperature/pressure dependence of sound velocities and lattice
parameters}
\author{H. V. Gomonay and V. M. Loktev}
\date{\today}
\affiliation{Bogolyubov Institute for Theoretical Physics NAS of
Ukraine,\\ Metrologichna str. 14-b, 03143, Kyiv, Ukraine}

\begin{abstract}In the present paper we generalised a phenomenological model developed in \cite{gomo:2005} for the
description of magnetostructural phase transitions and related
peculiarities of elastic properties in solid oxygen under high
pressure and/or low (below 40 K) temperature. We show that
variation of all the lattice parameters in the vicinity of
$\alpha\beta$-phase transition is due to both the shift of basal
closed packed planes and appearance of the long-range magnetic
order. Competition between these two factors from one side and
lattice compression below $T_{\alpha\beta}$ from another produces
non monotonic temperature dependence of lattice parameter $b$
(along monoclinic axis). Steep decrease of the sound velocities in
the vicinity of $T_{\alpha\beta}$ can be explained by the
softening of the lattice with respect to shift of the close-packed
planes (described by the constant $K_2$) prior to phase transition
point. We anticipate an analogous softening of sound velocities in
the vicinity of $\alpha \delta$-phase transition and non monotonic
pressure dependence of sound velocities in $\alpha$-phase.
\end{abstract}\date{\today} \maketitle
\section{Introduction}
Solid oxygen that belongs to a family of cryocrystals is being
studied for more than 100 years and still attracts attention of
researches. This simple molecular crystal has very complicated
temperature-pressure ($T-P$)phase diagram consisting of  magnetic
and nonmagnetic, metallic and dielectric, studied and
non-discovered phases (see, e.g., review \cite{Freiman:2004} and
references therein). Phase transitions and drastic change of the
magnetic, electronic, elastic properties of solid O$_2$ can be
triggered by variation of temperature, pressure, external magnetic
field, etc. The thorough analysis of corresponding regularities is
rather non-trivial problem because the O$_2$ crystal lattice is
mainly hold by the weak Van-der-Waals and so is very soft in
comparison with ordinary solids. Besides, in contrast to other
crystals, exchange magnetic interactions at low temperature prove
to be of  the same order as lattice energy. As a result,
application of external fields (temperature, stress, magnetic)
gives rise to a pronounce variation of all the lattice and
magnetic parameters. In this situation a precise microscopic
calculations should be combined with general thermodynamic
treatment.

In the I-st part of this paper \cite{gomo:2005} we made an attempt
to develop a phenomenologic (Landau-type) model aimed at the
description of the lattice and magnetic properties of solid oxygen
in a sequence of temperature/pressure  induced
$\beta\rightarrow\alpha\rightarrow\delta$ transitions.

The proposed model was based on the following assumptions:
\begin{enumerate}
\item Magnetoelastic coupling is so strong
that abrupt change of  the magnetic structure leads to the
noticeable variation of the crystal
  lattice.
  \item The primary order parameter in the series of $\beta\rightarrow\alpha\rightarrow\delta$ phase transitions
  is a homogeneous shift of closed-packed planes. The order
  parameter is defined with respect to the virtual hexagonal, D$_{6h}^1$
  pra-phase.
     \item\label{volume_assumption} The only macroscopic parameter that controls
     these
  transitions is the specific volume. Temperature dependence of
  the specific volume cannot be calculated within the model and
  should be
  taken from the experiment.
\end{enumerate}

As it was shown this model gives good agreement with experimental
data. However, some questions have been left beyond its scope. In
particular, we have considered only two factors that define
crystal structure of different phases -- shift of the closed
packed planes and specific volume of the crystal. More thorough
analysis should account for rather strong rhombic deformations
within the plane and separate contribution of interplane distance
and isotropic in-plane strain into the change of specific volume.

In the present paper we try to refine the model by taking into
account all the parameters that define the crystal structure of
$\beta$-, $\alpha$- and $\delta$- phases and explain observed
peculiarities of temperature/pressure dependence of lattice
constants and sound velocities.

We would also like to mention with our great pleasure that the
preliminary results of this paper were presented at International
Conference CC-2006 (in Kharkov) devoted to the prominent Ukrainian
scientist and wonderful woman Antonina Fedorovna Prihot'ko who is
well known for her brilliant experiments in physics of
cryocrystals. She is also famous for fundamental results in
optical investigation of $\alpha$ and $\beta$-phases in solid
oxygen. With the  present paper we try to pay our tribute of
respect to her memory.

\section{Model}
It was already mentioned that the crystal structure of $\beta$-,
$\alpha$- and $\delta$-phases can be considered as a homogeneously
deformed hexagonal lattice. In general, such a deformation can be
consistently described with the use of four independent variables:
lattice parameters $a$, $b$, $c$, and angle $\beta$ of monoclinic
cell. Corresponding combinations that form representation of the
D$_{6h}^1$ space symmetry group  (parent pra-phase) are:
\renewcommand{\theenumi}{\roman{enumi}}
\renewcommand{\labelenumi}{\theenumi)}
\begin{enumerate}
  \item relative
shift of closed-packed planes $\xi=(c/a)\cos\beta$;
  \item isotropic strain of basal plane $\Delta
  s/s_0=(ab-a_0b_0)/(a_0b_0)$;
  \item relative extension/contraction $u_{zz}=(h-h_0)/h_0$ in the direction
  perpendicular to closed-packed planes ($Z$-axis) , where
  $h\equiv c\cdot\sin\beta$;
  \item shear strain in the basal plane $u=(a-\sqrt{3}b)/(\sqrt{3}a_0b_0)^{1/2}$.
\end{enumerate}
The quantities with subscript ``0'' are attributed to a certain
reference state (at zero $T$ or $P$). In assumption of small
variation of interplane distance ( formally expressed by
inequality $u_{zz}\ll 1$) relative change of the specific volume
$\Delta v/v$ can be readily expressed in a form of a simple sum
$\Delta v/v=\Delta s/s_0+u_{zz}$.

Equilibrium values of $\xi$, $u_{zz}$, $u$, and $\Delta
  s/s_0$ at a given
temperature $T$ and hydrostatic pressure $P$ are calculated from
standard conditions for minimum of Gibbs' potential $\Phi$ that is
supposed to be invariant with respect to operations of the
symmetry group D$_{6h}^1$.

 Using results of symmetry analysis given in \cite{gomo:2005} (see
Table 2 therein) general expression for $\Phi$ can be represented
as a sum of structural, $\Phi_{\rm str}$, magnetic, $\Phi_{\rm
mag}$, elastic, $\Phi_{\rm elast}$ contributions and interaction
term $\Phi_{\rm int}$:
\begin{equation}\label{free-energy}
  \Phi=\Phi_{\rm str}+\Phi_{\rm mag}+\Phi_{\rm elast}+\Phi_{\rm int}.
\end{equation}
The structure of the first term in (\ref{free-energy})
\begin{eqnarray}\label{structural}
  \Phi_{\rm str}=-K_2\left(s, h\right)[\cos 2\pi\xi_1+\cos 2\pi\xi_2+\cos
  2\pi(\xi_1-\xi_2)]\nonumber\\
  +{1\over 4}K_4[\cos 4\pi\xi_1+\cos 4\pi\xi_2+\cos
  4\pi(\xi_1-\xi_2)]
  +P\left(\frac{\Delta s}{s}+u_{zz}\right)
\end{eqnarray}
reflects the translational invariance of the crystal with respect
to relative shift $\mathbf{u}$ of neighboring close-packed planes
of hexagonal lattice, namely:
$2\pi\xi_{1,2}\equiv\mathbf{b}_{1,2}\mathbf{u}$, where
$\mathbf{b}_{1,2}$ are vectors of the reciprocal pra-phase
lattice. Phenomenological constants $K_2$ and $K_4$ can be
considered as coefficients in Fourrier series of lattice
potential.

The coefficient $K_2$ is  supposed to be a linear function of the
isotropic in-plane strain and interplane distance with
corresponding phenomenological constants $\lambda_s,
\lambda_h\propto \lambda_v$ (where $\lambda_v= 10$~GPa was
estimated in \cite{gomo:2005}):
\[K_2(v)\equiv K_0-\lambda_s\frac{\Delta s}{s}-\lambda_hu_{zz}.\]
Both constants $\lambda_s, \lambda_h>0$ are supposed to have a
positive sign starting from general considerations (confirmed with
further calculations, see below). Namely, the effective constant
$K_2$ describes the strength of intermolecular forces that keep
relative arrangement of basal planes. Increase of the average
intermolecular distance should give rise to weakening of
intermolecular bonds and hence to a decrease of $K_2$.

Magnetic contribution in (\ref{free-energy})
\begin{eqnarray}\label{magnetic}
  \Phi_{\rm mag}=-J(\mathbf{k}_{13})\sum_{j=1}^2(\mathbf{l}^{(\beta)}_j)^2-J(\mathbf{k}_{12})\sum_{j=1}^3(\mathbf{l}^{(\alpha)}_j)^2-J(\mathbf{k}_{14})\sum_{j=1}^3(\mathbf{l}^{(\delta)}_j)^2,
\end{eqnarray}
accounts for the exchange interaction only, $J(\mathbf{k}_{j})>0$
are Fourier components of exchange integrals labeled according to
stars $\mathbf{k}_{j}$ ($j=12, 13, 14$) of irreducible
representations of D$_{6h}^1$ space group, antiferromagnetic (AFM)
vectors $\mathbf{l}^{(\beta)}$, $\mathbf{l}^{(\alpha)}$,
$\mathbf{l}^{(\delta)}$ unambiguously describe the magnetic
ordering in $\beta$-, $\alpha$- and $\delta$-phases\footnote{~We
assume that the series (\ref{free-energy}) may also include a wave
vector corresponding to 4-sublattice structure of recently decoded
$\epsilon$-phase \cite{Fujihisa:2006, Militzer:2006,
Lundegaard:2006}. Detailed analysis of this phase is out of scope
of that paper. }. It should be also stressed that the magnetic
ordering in $\beta$-phase has short-range nature (so called
correlation ordering with three sublattice 120$^\circ$ Loktev
structure \cite{Loktev:1975E, Freiman:2004, Vitebskii:1992}).
Collinear long-range magnetic ordering characterized by one of
$\mathbf{l}^{(\alpha)}$ vectors, is established in $\alpha$-phase.
An effective Neel temperature $T_N$ is close to the temperature
$T_{\alpha\beta}$ of $\alpha\beta$- transition
($T_{\alpha\beta}$=23.5~K and $T_N=40$~K at ambient pressure)
\cite{Freiman:2004}, so, saturation magnetization $M_0$ and,
correspondingly, $|\mathbf{l}^{(\alpha)}|=M_0$ noticeably depend
on temperature in the vicinity of $T_{\alpha\beta}$.

The elastic contribution in (\ref{free-energy})
\begin{equation}\label{nonlinear_elas}
  \Phi_{\rm elas}(\hat {u})=\frac{1}{2}c^\prime\left[({u}_{xx}-{u}_{yy})^2+4{u}_{xy}^2\right]+\frac{1}{2}c_{33}u_{zz}^2
\end{equation}
does not include strain tensor components $u_{xz}$, $u_{yz}$, that
can be considered as  small excitations over finite deformations
given by the order parameter $\xi_{1,2}$. An expression
(\ref{nonlinear_elas} ) is valid in the limit of infinitely small
strains $u_{xx}-u_{yy}$, $u_{zz}$. In such an  approximation
operations of plane shift  and deformation commutes, hence, all
the parameters of the model can be refer to the same ''zero''
state. Otherwise, one should specify the succession of lattice
transformations, use different reference frames  at each stage and
work within nonlinear elasticity approach.

Interaction energy includes terms that describe different
cross-over effects. Here we are concentrated on several of them.
First,  variation of interatomic  distances give rise to the
change of intra- (characterized with interaction constants
$\Lambda_{\rm intra}$, $\Lambda_{\perp}$), and inter- (constants
$\Lambda_{\rm inter}$, $\Lambda_{\|}$) plane exchange
integrals\footnote{~We keep superscripts $(\beta)$, $(\alpha)$ and
$(\delta)$ in the constants of magnetic nature only, in order to
emphasize step-like change of magnetic structure in the phase
transition point. Variation of other phenomenologic coefficients
is not so crucial.}. This effect has magnetoelastic origin and can
be formally described by two expressions. The first one is
responsible for isotropic
\begin{eqnarray}\label{interaction_m-e-is}
  \Phi^{(\rm iso)}_{\rm mag-el}&=&\left(\Lambda^{(\beta)}_{\rm intra}\frac{\Delta
  s}{s}+\Lambda^{(\beta)}_{\rm inter}u_{zz}\right)\sum_{j=1}^2(\mathbf{l}^{(\beta)}_j)^2\nonumber\\
  &+&\left(\Lambda^{(\alpha)}_{\rm intra}\frac{\Delta s}{s}+\Lambda^{(\alpha)}_{\rm inter}u_{zz}\right)\sum_{j=1}^3(\mathbf{l}^{(\alpha)}_j)^2\nonumber\\
  &+&\left(\Lambda^{(\delta)}_{\rm intra}\frac{\Delta s}{s}+\Lambda^{(\delta)}_{\rm inter}u_{zz}\right)\sum_{j=1}^3(\mathbf{l}^{(\delta)}_j)^2,
\end{eqnarray}
and another one for anisotropic effects
\begin{eqnarray}\label{interaction_m-e-an}
 \Phi^{(\rm an)}_{\rm mag-el}&=& -\Lambda^{(\alpha)}_{\|}\left[(\mathbf{l}^{(\alpha)}_1)^2\cos
2\pi\xi_1+(\mathbf{l}^{(\alpha)}_2)^2\cos
2\pi\xi_2+(\mathbf{l}^{(\alpha)}_3)^2\cos
  2\pi(\xi_1-\xi_2)\right]\nonumber\\
&-&\Lambda^{(\delta)}_{\|}\left[(\mathbf{l}^{(\delta)}_1)^2\cos
2\pi\xi_1+(\mathbf{l}^{(\delta)}_2)^2\cos
2\pi\xi_2+(\mathbf{l}^{(\delta)}_3)^2\cos
  2\pi(\xi_1-\xi_2)\right]\nonumber\\
  &-&\Lambda^{(\alpha)}_{\perp}\left\{(u_{xx}-u_{yy})(\mathbf{l}^{(\alpha)}_1)^2+2u_{xy}\left[(\mathbf{l}^{(\alpha)}_2)^2-(\mathbf{l}^{(\alpha)}_3)^2\right]\right\}\nonumber\\
  &-&\Lambda^{(\delta)}_{\perp}\left\{(u_{xx}-u_{yy})(\mathbf{l}^{(\delta)}_1)^2+2u_{xy}\left[(\mathbf{l}^{(\delta)}_2)^2-(\mathbf{l}^{(\delta)}_3)^2\right]\right\}.
\end{eqnarray}
Another coupling effect originates from strong anharmonocity which
is a peculiar feature of molecular crystals. Corresponding
contribution into thermodynamic potential has a form
\begin{eqnarray}\label{interaction_anhar-an}
 \Phi_{\rm anhar}&=& -\lambda_e\left\{(u_{xx}-u_{yy})\cos
2\pi\xi_1\right.\nonumber\\
 &+&2\left.u_{xy}\left[\cos 2\pi\xi_2-\cos
  2\pi(\xi_1-\xi_2)\right]\right\} -\lambda_\perp\frac{\Delta s}{s}u_{zz}.
\end{eqnarray}
where the coupling constants  $\lambda_e$, $\lambda_\perp$ are of
the same nature as $\lambda_s$, $\lambda_h$ but from the general
point of view should be much smaller in value (see Table
\ref{Table_1}).
\begin{table}
  \centering
  \caption{Phenomenological coefficients}\label{Table_1}
  \begin{tabular}{|c|c|c|}
    % after \\: \hline or \cline{col1-col2} \cline{col3-col4} ...
    \hline Coefficient& Value & Comments \\\hline
    $\chi_s$ & $0.31
\cdot 10^{-12}$cm$^2$/dyne
   &at amb. $P$,   \\
    &$3.2\cdot 10^{-12}$cm$^2$/dyne  & at $T$=19 K, $P=1\div 10$ GPa\\
    \hline
     $\beta_P(T)$& $1.6\cdot 10^{-4}$  1/K&  $T$<23.5 K  \\
  & $1.7\cdot 10^{-3}$  1/K&  23.5 K<$T$  \\\hline
 $K_{\rm eff}$&3
GPa& calc. at amb. $P$  \\\hline
     $K_4$&5 GPa&calc. at amb. $P$\\\hline
 $\lambda_s$& 13 GPa  &at amb. $P$,\\
     &0.2 GPa&at $T$=19 K, $P=1\div 10$ GPa\\\hline
      $\lambda_h$& 10 GPa & in assumption that $c_{33}$=10 GPa \\\hline
      $\lambda_\perp=c_{13}$& 1 GPa & in assumption that $c_{33}$=10 GPa \\\hline
       $\lambda_e$& 0.06 GPa  & in assumption that $c^\prime$=1 GPa  \\\hline
       $\Lambda^{(\alpha)}_{\|}M^2_0$&  -0.04 GPa&
\\\hline
$\Lambda^{(\alpha)}_{\perp}M^2_0$&  -0.02 GPa&
\\\hline
$\Lambda_{\rm {inter}}M^2_0$&  $\propto$ 0.06 GPa&in all phases
\\\hline
      \end{tabular}
\end{table}
 Really, $\lambda_s$, $\lambda_h$ describe
variation of average interatomic distances that result from
reconstruction of crystal lattice in the course of phase
transition. These constants are proportional to Grunaisen
parameter which is very high. At the same time $\lambda_e$ and
$\lambda_\perp$ are responsible for a ``differential'' effect,
namely, variation of anisotropy of the crystal lattice in the
course of phase transition.

Standard minimization procedure enables to obtain the value of
lattice parameters and stability conditions of $\beta-$, $\alpha-$
and $\delta-$ phases.

\section{Temperature/pressure dependence of lattice parameters}
 Equations for order parameter and stability
regions were derived and discussed in details in \cite{gomo:2005}.
The refinement of the model (splitting of in- and out-of-plane
contribution into specific volume) gives rise to no qualitative
changes. In this section we discuss only some additional effects
that can be explained and predicted in the framework of the model.
As in \cite{gomo:2005} we consider a homogeneous (from
crystallographic point of view) phase with
$\xi\equiv\xi_1=\xi_2/2$. An effective macroscopic order parameter
that vanishes in $\beta$-phase is introduced as $\eta\equiv
1+2\cos 2\pi\xi$. Magnetic ordering in $\alpha-$  and $\delta$-
phase is described by a single AFM vector
$\mathbf{l}_1^{(\alpha)}$  and $\mathbf{l}_1^{(\delta)}$,
correspondingly.
\subsection{Order parameter and isomorphic in-plane strain}
In the ``more symmetrical'' $\beta$- and $\delta$- phases an order
parameter takes the limiting values $\eta=0$ ($\beta$-phase) and
$\eta=-1$ ($\delta$-phase). Saturation magnetization $M_0$ (or
correlation parameter in $\beta$-phase) are constant. Pressure and
temperature dependence of the isomorphic in-plane strain $\Delta
s/s$ in these cases cannot be calculated from general
thermodynamic considerations and is determined experimentally as
\begin{equation}\label{surface beta}
\frac{\Delta s}{s}=\int_0^T\beta_s(T';P)dT'-\chi_s(T)P +
[\underbrace{\Lambda^{(\beta)}_{\rm intra}}_{\beta-\textrm{ph}}
+\underbrace{\Lambda^{(\delta)}_{\rm
intra}}_{\delta-\textrm{ph}}]M_0^2,
\end{equation}
where the in-plane thermal expansion coefficient $\beta_s(T;P)$
and isothermal $\chi_s(T)$ compressibility vary in a wide range
with pressure and temperature  (for example, Abramson et al
observed 10\% change of compressibility per 1 GPa in $\beta$-phase
at room temperature \cite{Abramson:1994}).

In $\alpha$-phase the temperature/pressure dependence of $\eta$
and $\Delta s/s$ can be calculated from the system
\begin{eqnarray}\label{minimum_conditions}
&&\left[K_0-K_4+\Lambda^{(\alpha)}_{\|}M_0^2-\lambda_s\frac{\Delta
s }{s}-\lambda_hu_{zz}\right]\eta
+\frac{1}{2}K_4(3\eta^2-\eta^3)-\Lambda^{(\alpha)}_{\|}M_0^2=0,\nonumber\\
&&\frac{\Delta
s}{s}=\int_0^T\beta_s(T';P)dT'-\chi_sP+\frac{\chi_s\lambda_s}{2}(\eta^2-3)-\chi_s\Lambda^{(\alpha)}_{\rm
intra}M_0^2(T).
\end{eqnarray}
(The behavior of interplane deformation $u_{zz}$ will be discussed
in the next section). The values of phenomenological constants
derived from fitting of functions $\eta(T,P)$, $\Delta s/s(T,P)$
to experimental data \cite{Krupskii:1979E, Akahama:2001} with due
account of $M_0(T)$ temperature dependence \cite{Bernabe:1997,
Bernabe:1998} are given in Table \ref{Table_1}. It can be seen
from the Table \ref{Table_1} that $K_{\rm
eff}=K_0-K_4+\Lambda^{(\alpha)}_{\|}M_0^2$ and $K_4$ are
comparable in order of value with the shear modulus of the
material \cite{Tarasenko:1981}. The value of $\chi_s$ which we
associate with the ``seed''compressibility is much smaller
compared with experimentally observed values \cite{Freiman:2004}.
This means that the main contribution into compressibility arises
from shifts of close-packed planes and magnetic interactions.

\subsection{Interplane distance}
Distance $h$ between the close-packed planes can be calculated
from equations
\begin{eqnarray}\label{interplane distance}
u_{zz}&=&-\frac{\lambda_\perp}{c_{33}}\frac{\Delta s}{s}
-\frac{\Lambda^{(\beta)}_{\rm inter}}{c_{33}}M_0^2\quad \textrm{for}\quad \beta-\textrm{phase},\nonumber\\
&=&-\frac{\lambda_\perp}{c_{33}}\frac{\Delta
s}{s}-\frac{\lambda_h}{2c_{33}}\eta^2
-\frac{\Lambda^{(\alpha)}_{\rm
inter}}{c_{33}}M_0^2(T)\quad \textrm{for}\quad \alpha-\textrm{phase},\nonumber\\
&=&-\frac{\lambda_\perp}{c_{33}}\frac{\Delta
s}{s}-\frac{\lambda_h}{2c_{33}} -\frac{\Lambda^{(\delta)}_{\rm
inter}}{c_{33}}M_0^2\quad \textrm{for}\quad \delta-\textrm{phase},
\end{eqnarray}
once the dependencies $\eta(T,P)$ and $\Delta s/s(T,P)$ are known.
Analysis of equations (\ref{interplane distance}) shows that in
the $\beta$-phase temperature/pressure dependence of $u_{zz}$ (and
hence interplane distance) is related with  lattice compression
within the close-packed plane. Thus, constant $\lambda_\perp$
coincides with the elastic modulus $c_{13}$. Estimated ratios
$c_{13}/c_{33}=0.1$ at ambient pressure and $c_{13}/c_{33}=0.24$
at $T=19 $~K, $P=1\div 7$~GPa and also the positive sign of
$c_{13}>0$ are in agreement with observations made by Abramson et
al. \cite{Abramson:1994}.

\begin{figure}[htbp]
\centering{ \epsfig{file=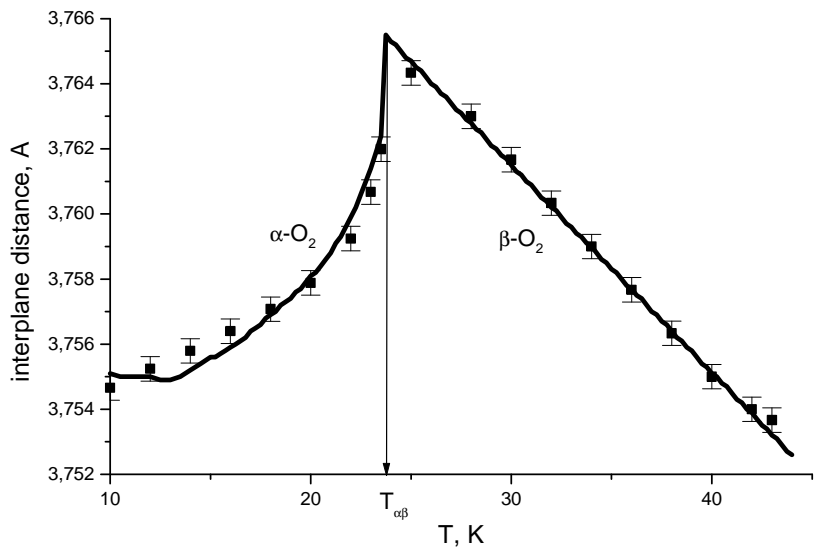, width=0.6\textwidth} }
\caption{Temperature dependence of interplane distance (squares)
calculated according to experimental data \cite{Krupskii:1979E}.
Theoretical curve (solid line) is calculated from Eq.
(\ref{interplane distance}). Arrow indicates the point of
$\alpha\beta$- transition.}\label{Fig_interplane_temp}
\end{figure}

\begin{figure}[htbp] \centering{ \epsfig{file=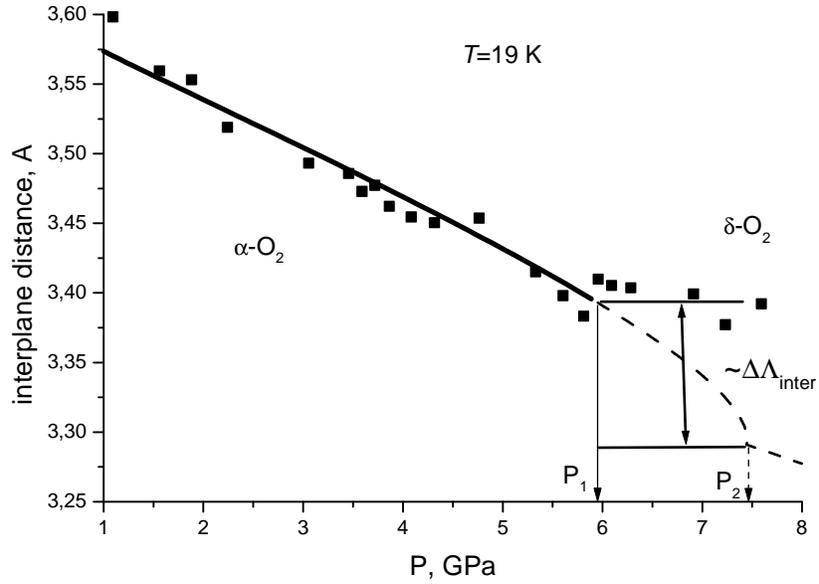,width=0.6\textwidth} }
\caption{Pressure dependence of interplane distance at $T$=19 K:
(squares) calculated according to experimental data
\cite{Akahama:2001}. Theoretical curve (solid line) is calculated
from Eq. (\ref{interplane distance}). Arrow indicates the point of
the real, $P_1$, and hypothetical, $P_2$, $\alpha\delta$-
transition.}\label{Fig_interplane_pressure}
\end{figure}

Figure \ref{Fig_interplane_temp}
 shows temperature
dependence of interplane distance experimentally measured
(squares) and calculated from (\ref{interplane distance}) (fitting
parameters calculated in assumption that $c_{33}=10$~GPa are given
in Table \ref{Table_1}).
Below $T_{\alpha\beta}$ an interplane distance abruptly diminishes
and then its temperature derivative changes sign. This fact may be
explained by the combined influence of the magnetic ordering
(characterized by difference $\Lambda^{(\beta)}_{\rm
inter}-\Lambda^{(\alpha)}_{\rm inter}$) and shift of basal planes
(corresponding interaction constant $\lambda_h$). In $\beta$-phase
each O$_2$ molecule is situated over the center of underlying
triangle. In-plane compression (that arises due to cooling  or
application of pressure) pushes out the overlaying molecules in
vertical direction, so interplane distance increases. In
$\alpha$-phase hexagonal planes are not only shifted from an ideal
''close-packed'' (with respect to underlying layer) position but
can also ``slip'' under compression. Appearance of an additional
degree of freedom allows lattice compression not only in-plane but
also in perpendicular direction. Formally this means that constant
$\lambda_h$ is positive, in agreement with monotonic decrease of
interplane distance under pressure-induced compression (see Fig.
\ref{Fig_interplane_pressure}).

Contribution of magnetic interactions into interplane distance is
important not only at $\alpha\beta$-, but also at $\alpha\delta$-
transition as can be seen from Fig. \ref{Fig_interplane_pressure}.
If we assume that $\alpha$- and $\delta$- phases have the same
magnetic structure, then pressure dependence of interplane
distance according to (\ref{interplane distance}) should be
continuous up to (hypothetical) 2-nd order transition point at
$P_2=7.5$~GPa (dashed line in Fig. \ref{Fig_interplane_pressure}).
Change of magnetic structure that originates from inter-plane
exchange interactions induces I-st order phase transition at
$P_1=6$~GPa$<P_2$ followed by abrupt change of the order parameter
$\eta$ and all the related parameters including $u_{zz}$. So,
difference between the observed and hypothetical $u_{zz}$ value
above $P_1$ (see Fig. \ref{Fig_interplane_pressure} and Eq.
(\ref{interplane distance})) is proportional to
$\Lambda^{(\delta)}_{\rm inter}-\Lambda^{(\alpha)}_{\rm inter}$
and is of magnetic nature.

\subsection{In-plane deformation: parameters $a$, $b$}
It is quite obvious from symmetry considerations that
establishment of long-range magnetic order in $\alpha$-phase is
followed by in-plane deformation described by $u_{xx}-u_{yy}\equiv
u$. Analysis of the expressions (\ref{interaction_m-e-an}),
(\ref{interaction_anhar-an}) shows that the same effect can be
produced by the shift of basal planes (term with $\lambda_e$) as
seen from the following equation:
\begin{equation}\label{shear_strain}
  u=\frac{\lambda_e}{c^\prime}(\eta^2-3\eta)+\frac{\Lambda_\perp^{(\alpha)}M_0^2(T)}{c^\prime}.
\end{equation}

\begin{figure}[htbp] \centering{ \epsfig{file=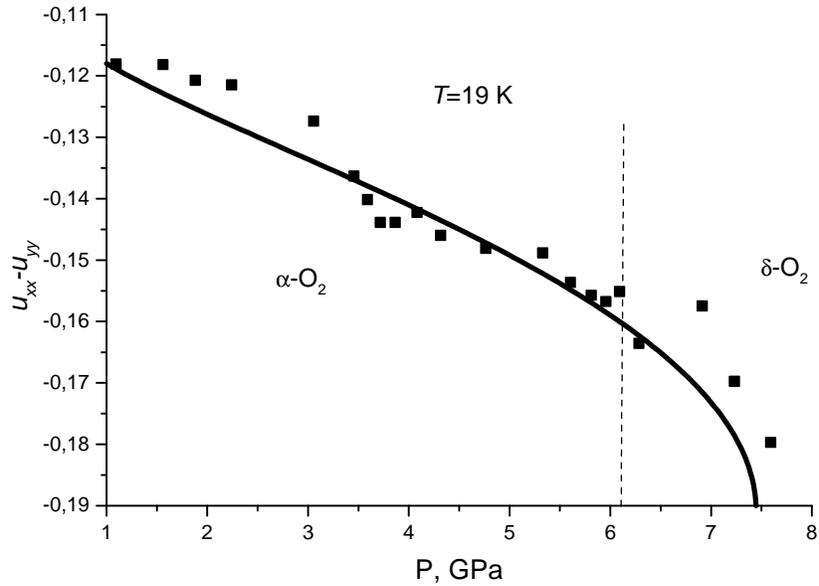,width=0.6\textwidth} }
\caption{Pressure dependence of in-plane rhombic strain
$u_{xx}-u_{yy}$ (solid line) calculated from
Eq.(\ref{shear_strain}) . Squares: experimental data
\cite{Akahama:2001}.}\label{Fig_u_pressure}\end{figure}

\begin{figure}[htbp] \centering{ \epsfig{file=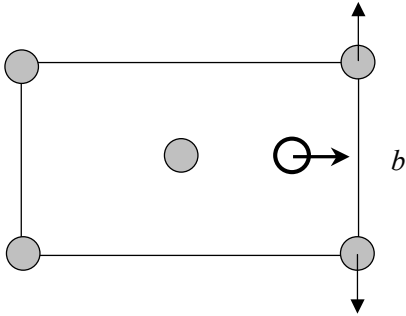,width=0.3\textwidth} }
\caption{In-plane displacements (arrows) of O$_2$ molecules
induced by relative shift of the overlaying (open circle)
plane.}\label{Fig_shift}\end{figure}

Pressure dependence of $u_{xx}-u_{yy}$ calculated from Eq.
(\ref{shear_strain}) along with experimental data is given in Fig.
\ref{Fig_u_pressure}.
 It is
clearly seen that in $\alpha$-phase lattice anisotropy
$u_{xx}-u_{yy}$ monotonically increases in absolute value with
pressure. This effect can be explained from very simple
considerations. Pressure produces compression in all directions in
the basal plane but relative shift of planes can induce an
effective tension in $b$ direction as seen from
Fig.\ref{Fig_shift}  (O$_2$ molecules move apart in $b$ direction
when overlaying molecule moves toward the edge). As a result
deformation in basal plane is essentially anisotropic and
anisotropy is more pronounced at smaller interatomic distances
(i.e., at higher pressure).

\begin{figure}[htbp]
\centering{ \epsfig{file=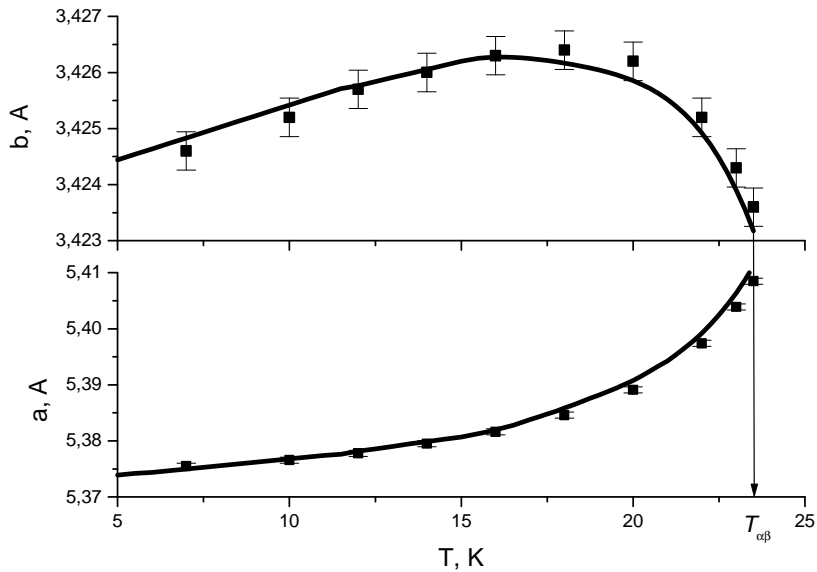,width=0.6\textwidth} }
\caption{Temperature dependence of lattice parameters $a$ and $b$
(solid lines) calculated from Eq., (\ref{minimum_conditions}),
(\ref{shear_strain}). Squares: experimental data
\cite{Krupskii:1979E}).}\label{Fig_ab_temeprature}
\end{figure}

The same effect takes place during cooling $\alpha$-phase at
ambient pressure. Fig. \ref{Fig_ab_temeprature}  shows temperature
dependence of  $a$ and $b$ lattice parameters below
$T_{\alpha\beta}$. Intermolecular lattice distance $a$ decreases
monotonically because of cooling-induced compression. In contrast,
temperature dependence of the parameter $b$ is non-monotonic.
Increase of $b$ during cooling from 23.5 to 18 K means that
tension in this direction is stronger than cooling-induced
compression. In particular, Eq. (\ref{shear_strain}) demonstrates
that two mechanisms may be responsible for this behaviour:
repulsion of ferromagnetically ordered neighbors
\cite{Freiman:2004} (described by the term
$\Lambda_\perp^{(\alpha)}M_0^2(T)$, $M_0^2$ increases with
temperature decrease)  and already mentioned shift of basal
planes.

Comparison of the above theoretical dependences with experimental
data obtained by different groups using different technique makes
it possible to estimate the values and range of phenomenological
constants (see Table \ref{Table_1}). All the coefficients could be
grouped in three categories: i) large ($\propto 10$~GPa) constants
responsible for anharmonicity effects ($\lambda_{s,h}$); ii)
intermediate ($\propto 1$~GPa) constants ($K_{\rm eff}$, $K_4$,
$c_{13}$) that characterize elastic properties of crystal and iii)
small (below 0.1~GPa) constants that describe magnetic
interactions and anisotropy effects (rest of the constants).
Though small in value the magnetic interactions reveal themselves
in the case of competition between different structural
interactions like pressure induced compression and repulsion that
arises due the shift of basal planes. Strong anharmonicity is
quite natural phenomena for molecular crystals. It is remarkable
that at ambient pressure both coefficients $\lambda_{s}$ and
$\lambda_{h}$ are of the same order and the main contribution into
effective structural constant $K_2$ arises from isomorphic
in-plane strain\footnote{~We are grateful to Prof. Yu. Freiman who
has drawn our attention to this fact.} $\Delta s/s$.

\section{Peculiarities of sound velocities in the vicinity of phase transition points}
Experimental study of sound velocity in solid oxygen provides
information about interactions that play the leading role in phase
transitions. High pressure measurements in $\beta$-phase
\cite{Abramson:1994} show monotonic increase with pressure of all
the elastic constants except shear modulus $c_{44}$, the value of
which abruptly decreases in the vicinity of
$\beta\delta$-transition point (approximately 8~GPa at 295~K).
Low-temperature curves \cite{Tarasenko:1969, Tarasenko:1981} have
a pronounced minima at $T_{\alpha\beta}$.

Critical behaviour of sound velocities can be quite naturally
explained in the framework of the developed model using the
concept of Goldstone mode. Really,the II-nd (or weak
I-st\footnote{~In other words, I-st order phase transition between
the phases that are in subgroup relation, i.e. I-st order close to
II-nd one.}) order phase transitions are  usually accompanied with
softening of a certain bond responsible for appearance of the
order parameter. So, characteristic frequency of the corresponding
excitations (symmetry related to the order parameter) should
vanish or at least noticeably diminish in the vicinity of the
phase transition point (Goldstone mode). In the case of
$\beta\rightarrow\alpha\rightarrow\delta$ transitions an order
parameter (shift of basal planes) is symmetry related with strain
tensor component $u_{zx}$ (see Table 2 in \cite{gomo:2005}) and,
as a result, with the transverse acoustic modes propagating in
[0001] and [1000] directions. Thus, these modes should have
peculiarity in phase transition points.

Sound velocities of the ``soft'' transverse, $v_t$, and
longitudinal, $v_l$ acoustic waves propagating in [1000] direction
are expressed through the elastic modula in a standard way
\begin{eqnarray}\label{velocity}
v^2_{t}&=&\frac{1}{2\rho}\left[c_{11}+c_{55}-\sqrt{(c_{11}-c_{55})^2+4c^2_{15}}\right],\nonumber\\
v^2_l&=&\frac{1}{2\rho}\left[c_{11}+c_{55}+\sqrt{(c_{11}-c_{55})^2+4c^2_{15}}\right],
\end{eqnarray}
where $\rho$ is crystal density.

Elastic modula that equal to the 2-nd derivatives of thermodynamic
potential $\Phi$ (see Eq.(\ref{free-energy})) with respect to
strain tensor components are calculated at equilibrium values of
lattice parameters:
\begin{eqnarray}\label{modulus}
c_{11}=\left.\frac{\partial^2 \Phi}{\partial (\Delta
s/s)^2}\right|_0,\quad
c_{55}=\left(\frac{c\sin\beta}{a}\right)^2\left.\frac{\partial^2
\Phi}{\partial \xi^2}\right|_0,\nonumber\\
c_{15}=\frac{c\sin\beta}{a}\left.\frac{\partial^2 \Phi}{\partial
\xi\partial (\Delta s/s)}\right|_0,\quad
du_{xz}=\left(\frac{c\sin\beta}{a}\right)d\xi.
\end{eqnarray}
Factor $(c/a)\sin\beta$ in (\ref{modulus}) is deduced from
geometrical considerations.

Then, omitting some terms immaterial for further discussion we
obtain that in all three phases $c_{11}=\chi_s^{-1}$, $c_{15}$=0
in $\beta$- and $\delta$-phases and
\begin{equation}\label{c55beta}
  c_{55}=\left(\frac{4\pi
  c\sin\beta}{a}\right)^2\left[\lambda_s(\beta_sT-\chi_sP)-\frac{3}{2}K_{\rm
  eff}\right]
\end{equation}
in $\beta$- phase,
\begin{equation}\label{c55delta}
  c_{55}=\left(\frac{4\pi
  c\sin\beta}{a}\right)^2\left[\lambda_s(\chi_sP-\beta_sT)+K_{\rm
  eff}-K_4+2\Lambda^{(\alpha)}_{\|}M_0^2\right]
\end{equation}
in $\delta$-phase,
 and
\begin{eqnarray}\label{c55alpha}
  c_{55}&=&\left(\frac{4\pi
  c\sin\beta}{a}\right)^2\left[\lambda_s(\chi_sP-\beta_sT)-K_{\rm
  eff}-3K_4\eta\left(1-\frac{\eta}{2}\right)\right]\sin^22\pi\xi,\nonumber\\
  c_{15}&=&\frac{4\pi
  c\sin\beta}{a}\lambda_s\sin2\pi\xi
\end{eqnarray}
in $\alpha$-phase.

It can be easily seen from (\ref{c55beta}) that in $\beta$-phase
shear modulus increases with temperature and decreases with
pressure, at least in the vicinity of transition point, in
consistency with observations \cite{Abramson:1994}.

\begin{figure}[htbp] \centering{ \epsfig{file=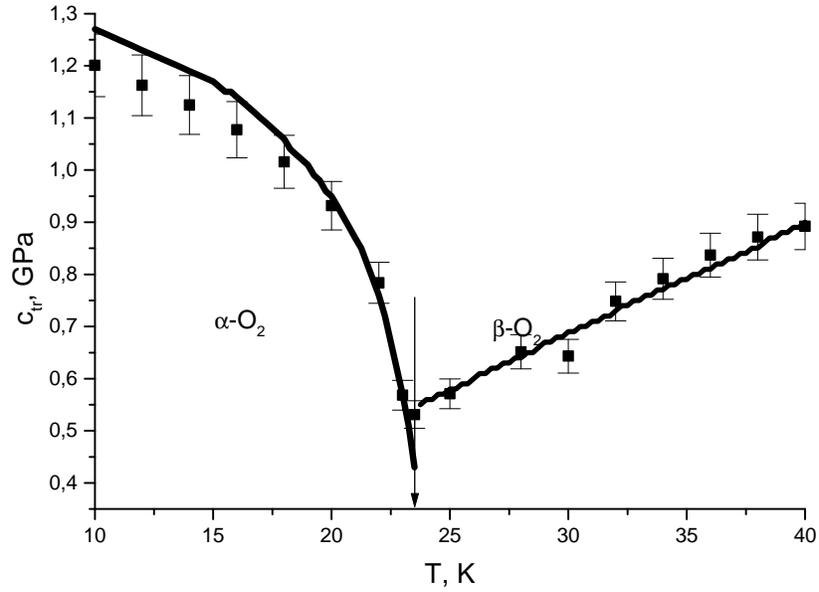,width=0.6\textwidth} }
\caption{Temperature dependence of shear modulus $c_t$ calculated
from (\ref{velocity})-(\ref{c55alpha}) (solid line). Squares:
experimental data
\cite{Tarasenko:1981}.}\label{Fig_transverse_temperature}\end{figure}

\begin{figure}[htbp] \centering{ \epsfig{file=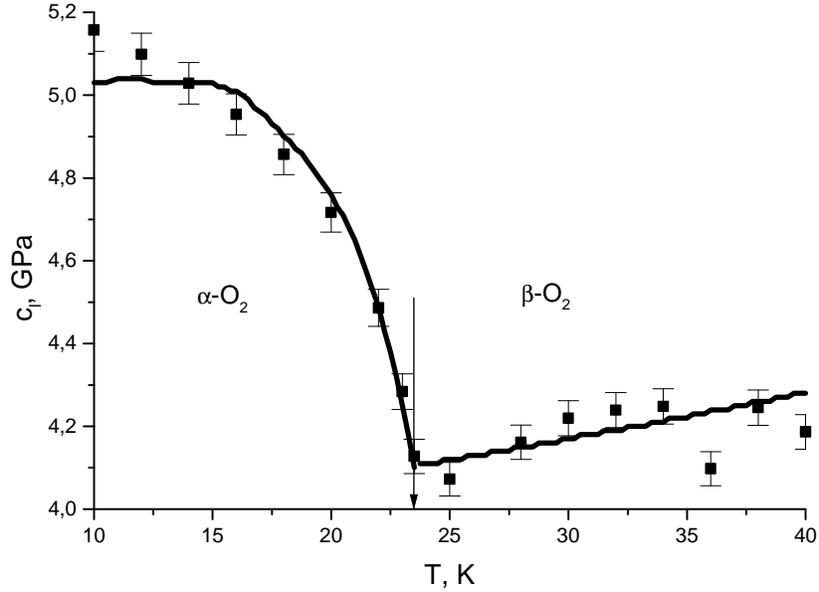,width=0.6\textwidth} }
\caption{Temperature dependence of compression modulus $c_l$
calculated from (\ref{velocity})-(\ref{c55alpha})  (solid line).
Squares: experimental data
\cite{Tarasenko:1981}.}\label{Fig_logitudinal_t}\end{figure}

More interesting is comparison of experimental and theoretical
temperature dependencies\footnote{~Temperature dependence of
$c_{11}=\chi^{-1}_s$ is taken from fitting the data in
$\beta$-phase.} $c_{t,l}=\rho v_{t,l}^2$ shown in
Figs.\ref{Fig_transverse_temperature}, \ref{Fig_logitudinal_t}.
Anomalous softening of $c_t$ agrees well with theoretical
predictions and thus may be explained by high compliance of O$_2$
crystal lattice with respect to shift of the closed-packed
molecular planes. Softening of $c_l$ is not so obvious from
general point of view but in the case of molecular crystal may
originate from strong anharmonicity.

\begin{figure}[htbp] \centering{ \epsfig{file=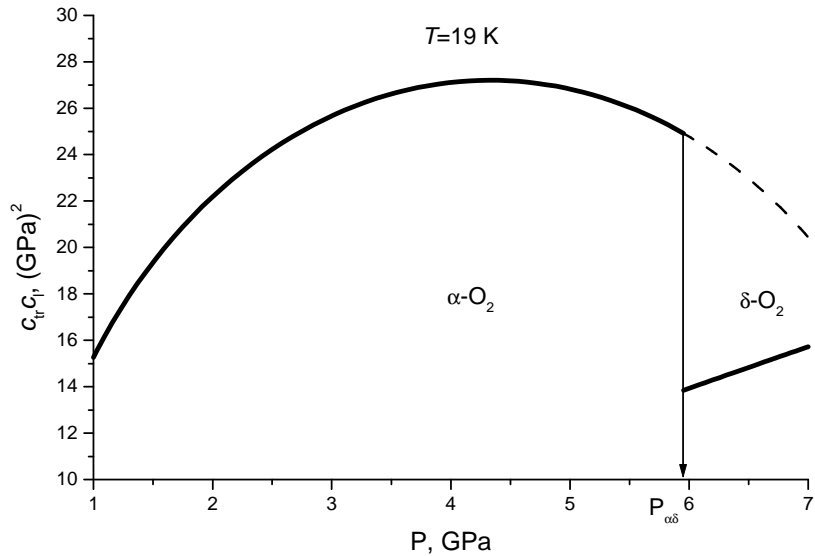,width=0.6\textwidth} }
\caption{Pressure dependence of the product of $c_tc_l$ calculated
from (\ref{velocity})-(\ref{c55alpha})  (solid
line).}\label{Fig_transverse_pressure}\end{figure}

 An
analogous softening of elastic modula and corresponding sound
velocities is also expected in the vicinity of
$\alpha\delta$-transition. Fig. \ref{Fig_transverse_pressure}
shows hypotetical pressure dependence of the modula
product\footnote{~The lack of experimental data makes it
impossible to separate $c_t$  and $c_l$ contributions.} $c_tc_l$
calculated on the basis of equations
(\ref{velocity})-(\ref{c55alpha}) with characteristic values of
phenomenologic parameters (see Table \ref{Table_1}). It is clear
that this dependence  should be nonmonotonic, with noticeable
decrease of modulus while approaching to $\alpha\beta$ and
$\alpha\delta$-phase boundaries.
\section{Conclusions}
In summary, we have calculated temperature and pressure dependence
of all the crystal lattice parameters of solid oxygen in the
magnetic $\alpha\beta\delta$ phases. The above phenomenological
model ascribes the leading role in structural changes to the shift
of satisfactory  agreement with the experiment.

In particular, shift of the planes is responsible for nonmonotonic
temperature dependence of lattice parameter $b$ in $\alpha$-phase
(effective repulsion), change of the effective thermal coefficient
in $c$ direction in $\alpha\beta$ transition (nonmonotonic $t$
dependence of  interplane distance), softening of shear modulus
and corresponding velocities in the vicinity of $T_{\alpha\beta}$.

We assume that with certain stipulation the developed model may be
applied to interpretation of IR spectra \cite{Santoro:2001,
Kreutz:2005} which pressure dependence looks similar to pressure
dependence of the order parameter. In particular, we predict
non-monotonic pressure dependence of elastic modula in
$\alpha$-phase and critical behaviour in $\alpha\delta$-
transition point. At the same time some of the problems are still
being outside of theoretical treatment. First of all it concerns
magnetocrystalline structure of $\epsilon$-phase, metallization of
highly compressed oxygen along with corresponding mechanism of
superconductivity, optical properties of high pressure phases,
etc. These and other problems are the subject of future
investigations.

We would like to acknowledge all the participants of CC-2006
Conference for keen interest to our presentations and fruitful
discussions.
\flushleft                      % библиографию не выравнивать
%\bibliographystyle{unsrt}  % стиль в духе ГОСТа
%\bibliography{oxygen_biblio}

\end{document}